\documentclass[aps,prd,reprint,nofootinbib,longbibliography,showkeys]{revtex4-2}

\pdfoutput=1

\usepackage{blindtext}
\usepackage{mathtools}
\usepackage{xcolor}

\usepackage{dsfont}
\usepackage{siunitx}

\usepackage{mathrsfs}
\usepackage{braket}

\usepackage[utf8]{inputenc}
\usepackage[english]{babel}
\usepackage{hyperref}

\hypersetup{
    colorlinks=true,
    linkcolor=blue,
    filecolor=blue,      
    urlcolor=blue,
    citecolor=blue,
}
\usepackage{color,colortbl}

\usepackage{tensind}
\tensordelimiter{?}

\usepackage{graphicx}
\usepackage{hyperref}
\DeclareGraphicsExtensions{.bmp,.png,.jpg,.pdf}
\usepackage{amsmath}
\usepackage{courier}
\usepackage{verbatim}
\usepackage{amssymb}
\usepackage{amsfonts}
\usepackage{natbib}
\usepackage{caption}
\usepackage{subcaption}

\usepackage{booktabs}
\usepackage{siunitx}

\newcommand{\be}{\begin{equation}}
\newcommand{\ee}{\end{equation}}
\begin{document}
\title{Single field matter bounce with dark energy era: comparison with CMB Planck 2018 data and best fit parameters}

\author{Rodrigo F. Pinheiro$^{1}$, Nelson Pinto-Neto$^{1}$}
\email{rfpinheiro@cbpf.br}
\email{nelsonpn@cbpf.br}

\affiliation{$^1$ Centro Brasileiro de Pesquisas F\'{\i}sicas - CBPF, Rua Dr. Xavier Sigaud 150, Urca, 22290-180, Rio de Janeiro, RJ, Brazil}

\begin{abstract}
In this work, we perform Markov Chain Monte Carlo (MCMC) analyses using the Planck 2018 cosmic microwave background (CMB) datasets, including temperature, polarization, and lensing, in order to compare matter bounce models with observational data. The particular model we considered contains a scalar field with an exponential potential, which behaves as dust in the asymptotic past of the contracting phase, it realizes a quantum bounce, and then behaves as a transient dark energy field at large scales in the expanding phase. The parameter $\lambda$ appearing in the exponential potential is directly related to the model's scalar spectral index, $n_s$, which is set free in the MCMC analyses, as well as the deepness of the bounce, which controls the amplitude of the power spectrum. We provide constraints on the cosmological parameters and compare the model's performance against the standard inflationary $\Lambda$CDM scenario. Our results indicate that Planck data alone cannot favor one model with respect to the other, showing that the model we investigate can be a viable alternative to inflation. 
\end{abstract}

\maketitle

\section{Introduction}

The primordial power spectrum is fundamental to establish the initial conditions for the distribution of the primary components of the universe: baryons, radiation, and dark matter. The Cosmic Microwave Background (CMB) originates from the complex interactions between baryons and photons within the evolving gravitational field, spanning from the era of recombination to the decoupling of matter and radiation \cite{Mukhanov:2005sc, Dodelson:2020bqr}. Diverse models of the early universe propose distinct mechanisms for the origin of these initial conditions. In the inflationary paradigm, fluctuations entering the Hubble radius during the subsequent radiation-dominated era had established causal contact during the inflationary phase, explaining the near temperature isotropy of the CMB \cite{Linde:2007fr, Baumann:2009ds, Achucarro:2022qrl}. Also, in bouncing models, modes have sufficient time to interact and thermalize during the contracting phase. Hence, when they cross the Hubble radius during the expanding phase, they possess a prior causal connection \cite{Pinto-Neto:2021gcl, Pinto-Neto:2021jko, Peter:2008qz}. 

Furthermore, the features of small linear temperature anisotropies in the CMB (amplitude and spectral index) can also be explained by a large class of slow-roll inflationary models, and some bouncing models with matter dominated contraction \cite{Wands:1998yp, Finelli:2001sr}. Increasingly precise measurements of the CMB are essential for discriminating between these competing early-universe scenarios. A critical first step involves constraining these models against the Planck 2018 data \cite{Planck:2019nip} using two-point statistics. Ensuring consistency at the two-point level provides the necessary justification for more rigorous investigations utilizing higher-order statistics, such as the bispectrum or trispectrum.

In Ref.~\cite{Bacalhau:2017hja} it was presented a bouncing model in which the contracting phase is dominated by a canonical scalar field with exponential potential, which behaves as an almost pressureless fluid in the far past of this phase. After the bounce, when the usual matter components of the standard cosmological model are supposed to be created,\footnote{The bounce period should play the role of the reheating phase after inflation.} the universe enters into the usual standard model expanding phases, when the scalar field becomes sub-dominant, returning to be important only when it behaves as a transient dark energy component. It was shown that the primordial scalar and tensor perturbations arising from this early matter bounce contraction can have the right observed amplitudes and spectral indices, depending on the deepness of the bounce and a parameter appearing in the scalar field exponential potential.

In the present paper, we confront this specific bouncing model with Planck data. Besides the usual free parameters of the standard cosmological model (baryon density $\Omega_{b} h^{2}$, cold dark matter density $\Omega_{c} h^{2}$, the present Hubble parameter $H_0$, and the optical depth $\tau$), we add the two other free parameters mentioned above: the deepness of the bounce, and the scalar field potential parameter. In order to be as general as possible when comparing the above bouncing model with inflationary models, we just impose that inflationary scenarios generically lead to power spectrum of scalar perturbations given by $P_k=A_s k^{n_s-1}$, with $A_s \approx 10^{-10}$ and  $n_s \approx 1$, knowing that such parameters are usually connected with the energy scale of inflation and slow-roll parameters. We show that not only the bouncing model of Ref.~\cite{Bacalhau:2017hja} can be well fitted with Planck data, but also that it is indistinguishable from inflation at linear level using Planck data alone.

The paper is divided as follows; in the next section we describe the background model, in Sec.~III we present the perturbation analysis, in Sec.~IV we describe the methodology used do constrain the parameter space of both type of models, in Sec.~V we present our results, and we end up with the conclusions in Sec.~VI. In the Appendix we present the details of the matching procedures to connect the quantum background phase to its classical evolution.

\section{Background}

The present cosmological model has a scalar field coupled to the metric space-time, with exponential potential $V(\phi) = V_{0} e^{-\lambda \kappa\phi}$, where $\kappa \equiv 1/M_{p} \equiv \sqrt{8 \pi G_{N}}$, $G_{N}$ is the Newton constant, $\lambda$ and $V_{0}$ are free parameters. The equations of motion can be derived from the following Lagrangian density

\begin{equation}
\mathcal{L} = \sqrt{-g}[\nabla^{\nu}\nabla_{\nu}\phi - V(\phi)].  
\end{equation}

The model space-time is flat, homogeneous and isotropic, described by the Friedmann-Lamâitre-Robertson-Walker metric    

\begin{equation}
    ds^2 = N^2(\tau)d \tau^2 - a(\tau)^2(dx^2 + dy^2 + dz^2),
\end{equation}
where $N^2(\tau)$ is the lapse function and $a(\tau)$ is the scale factor. When $N(\tau) = 1 \rightarrow \tau = t$, we have cosmic time. The equation of motion is the Friedmann equation
\begin{equation}
\label{eq:friedmann}
  \dot{H} = -\frac{\kappa^2}{2} \dot{\phi}^2, 
\end{equation}
coupled to the Klein-Gordon equation
\begin{equation}
\label{eq:klein-gordon}
    \ddot{\phi} = -3H\dot{\phi} - \frac{dV}{d \phi}.
\end{equation}
The dots represent the cosmic time derivative, the Hubble function is defined by $H = \dot{a}/a$ and it must satisfy the Friedmann constraint

\begin{equation}
    H^2 = \frac{\kappa^2}{3}\left[ \frac{\dot{\phi}^2}{2} + V(\phi)\right].
\end{equation}

We can write the background equations in a simple form, choosing the following dimensionless variables

\begin{equation}
    x = \frac{\kappa \dot{\phi}}{\sqrt{6} H}, \quad y = \frac{\kappa \sqrt{V}}{\sqrt{3} H}.
    \label{eq:x-y}
\end{equation}

Using these variables, the Friedmann constraint and the equation of state can be written as
\begin{equation}
\label{eq:friedmann-planar}
    x^2 + y^2 = 1, \quad \omega = 2x^2 - 1.
\end{equation}
When $x = \pm 1 \rightarrow \omega = 1 $,  the scalar field behaves as stiff-matter, and when $x = 0 \rightarrow \omega  = -1$, it behaves as dark energy. 

The Friedmann equation \eqref{eq:friedmann} and the Klein-Gordon equation \eqref{eq:klein-gordon} are rewritten as the following planar system  
\begin{equation}
\label{eq:planar-1}
    \frac{dx}{d \alpha} = -3x(1 - x^2) + \lambda \sqrt{\frac{3}{2}}y^2,
\end{equation}
    
\begin{equation}
\label{eq:planar-2}
    \frac{dy}{d \alpha} = xy\left(3x - \lambda \sqrt{\frac{3}{2}}\right),
\end{equation}
with $\dot{\alpha} = H$ and $\dot{H} = -3H^2x^2$ as auxiliary equations. It can be shown from this last equation that when $x \rightarrow \pm 1$, $H \rightarrow -\infty $, in the contracting phase, and $H \rightarrow +\infty $, in the expanding phase. These are singular points in the classical framework.   

Combining the Friendmann constraint \eqref{eq:friedmann-planar} with the first planar equation \eqref{eq:planar-1} we can have 

\begin{equation}
    \frac{dx}{d\alpha} = -3\left(x - \frac{\lambda}{\sqrt{6}}\right)(1 - x)(1 + x).
    \label{eq:classical-equation}
\end{equation}
The critical points of the system are $x = \pm 1$ and $x = \lambda/\sqrt{6}$. A study of stability around these points can demonstrate that $x\pm 1, y = 0$ are repeller points in the expanding phase, and attractor points in the contracting phase, and we have the opposite behavior for $x = \lambda/\sqrt{6}$, i.e, it is an attractor point in the expanding phase, and a repeller point in the contracting phase.   

Choosing the initial condition for $x$, as $x_{\lambda} = \frac{\lambda}{\sqrt{6}} + \epsilon$, where $0 < \epsilon \ll 1$, leads to a contracting phase until close to the singularity, when quantum effects take place - which we will address in the next subsection - launching the universe to an expanding era, passing through a dark energy phase, when $x = 0$, ending up in the attractor point. If $\lambda = \sqrt{3}$, the universe begins and ends in a matter fluid epoch. This choice leads to a scale invariant primordial power spectrum, but in this work we treat $\lambda$ as a free parameter in our Markov Chain Monte Carlo (MCMC) analyses, determining its best-fit value using Planck 2018 Cosmic Microwave Background (CMB) data. 

\begin{figure}
    \centering
    \includegraphics[width=0.5\textwidth]{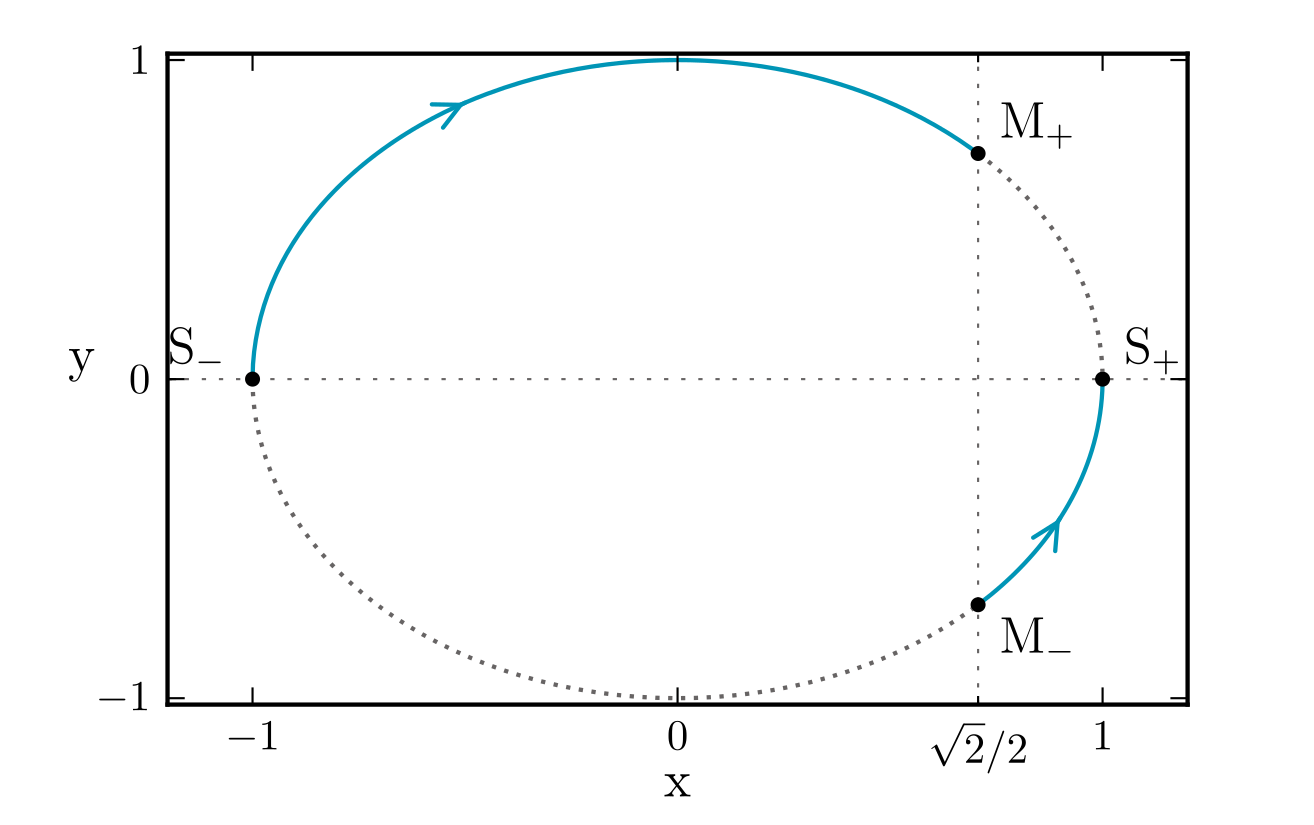}
    \caption{The planar system with $\lambda = \sqrt{3}$ and dark energy era in the expanding phase. The system starts at the repeller point $M_{-}$, in the contracting phase ($y < 0$), and ends up at the attractor point $M_{+}$, in the expanding phase ($y>0$). Figure taken from reference \cite{Bacalhau:2017hja}.}
    \label{fig:phase_space_B}
\end{figure}

The evolution of the planar system with $\lambda = \sqrt{3}$ and a dark energy era in the expanding phase can be seen in the figure \ref{fig:phase_space_B}. $M_{-}$ and $M_{+}$ label the points $x_{\lambda}$ in the contracting and expanding phases, respectively.  $S_{-}$ and $S_{+}$ are the singular points in $y = 0$ and $x \mp 1$, respectively. 

\subsection{Quantum bounce}

The evolution of the universe in the contraction phase leads it to a regime of extremely high energies, where the classical theory of general relativity fails. In this high-density regime, it becomes necessary to combine quantum mechanics with gravitation. Here, as we are restricted to homogeneous geometries (minisuperspace), we adopt the Wheeler-DeWitt quantization \cite{DeWitt:1967ub}, which is assumed to be a good approximation if the model scales never gets to close to the Planck scale. 

Close to the maximum contraction, the kinetic term is dominant. Changing to a dimensionless scalar field $\phi \to \kappa \phi/\sqrt{6}$, we can write the Hamiltonian that governs the system dynamics as

\begin{equation}
    H = N \mathcal{H} = \frac{N \kappa^2}{2 \ell_{p} e^{3\alpha}}(-\Pi^2_{\alpha} + \Pi^2_{\phi}),
\end{equation}
where $\ell_{p} \equiv \sqrt{G_{N}}$ is the Planck lenght, $\Pi_{\alpha}$ and $\Pi_{\phi}$ are the canonically conjugate momenta to $\alpha$ and $\phi$, respectively:

\begin{equation}
    \Pi_{\alpha} = -\frac{\ell_{p}}{N } e^{3 \alpha}\frac{ d\alpha}{d \tau}, \quad \Pi_{\phi} = \frac{\ell_{p}}{N} e^{3 \alpha}\frac{d \alpha}{d \tau}.
\end{equation}

General relativity is invariant through time re-parametrization, which implies that the Hamiltonian is constrained to vanish. Performing the Dirac quantization procedure, the canonical variables $\alpha$, $\phi$, $\Pi_{\alpha}$ and $\Pi_{\phi}$ are promoted to quantum operators, also the Hamiltonian constraint must annihilate the wave function, this condition leads to the Wheeler-DeWitt equation    

\begin{equation}
    \hat{\mathcal{H}}\Psi(\alpha, \phi) = 0 \Rightarrow \left[-\frac{\partial^2}{\partial\alpha^2} + \frac{\partial^2}{\partial\phi^2}\right]\Psi(\alpha, \phi) = 0.
\end{equation}

This is a Klein-Gordon-like equation in minisuperspace, whose general solution is a superposition of plane waves

\begin{align}
    \nonumber \Psi(\alpha, \phi) &= F(\phi + \alpha) + G(\phi - \alpha)\\
     \nonumber &\equiv \int dk \{f(k)\exp[ik(\phi + \alpha)] +\\
      &+ g(k)\exp[ik(\phi - \alpha)]\}.
\end{align}
The functions $f$ and $g$ were chosen to allow Gaussian superposition:
\begin{equation}
    f(k) = g(k) = \exp\left[\frac{-(k - d)^2}{\sigma^2}\right].
\end{equation}
where $d$ and $\sigma$ are parameters emerging from the initial conditions imposed to the wave function of the universe. Under this prescription, the wave function takes the form,

\begin{align}
    \nonumber \Psi(\alpha, \phi) &= \sigma \sqrt{\pi}\left\{\exp{\left[-\frac{(\alpha + \phi)^2 \sigma^2}{4}\right]} \right\}\exp{[id(\alpha + \phi)]} \\
      &+ \exp{\left[-\frac{(\alpha - \phi)^2\sigma^2}{4}\right]}\exp{[-id(\alpha - \phi)]}. \label{eq:wave-function}
\end{align}

Using the de Broglie-Bohm (dBB) quantum theory \cite{bohm1952suggested, holland1995quantum, durr2009bohmian}, quantum trajectories for $\alpha(t)$ and $\phi(t)$ can be calculated through the dBB guidance relations in cosmic time $N = 1$,

\begin{equation}
\label{eq:dBB1}
    \Pi_{\alpha} = \frac{\partial S}{\partial \alpha} = - \ell_{P} e^{3\alpha}\dot{\alpha},
\end{equation}

\begin{equation}
\label{eq:dBB2}
    \Pi_{\phi} = \frac{\partial S}{\partial \phi} = - \ell_{P} e^{3\alpha}\dot{\phi}.
\end{equation}

Taking the phase $S$ from the wave function \eqref{eq:wave-function} and substituting the result into the dBB guidance relations \eqref{eq:dBB1}, \eqref{eq:dBB2}, we arrive at the following planar system that describes the Bohmian trajectories:  

\begin{align}
    \ell_{P}\dot{\alpha} &= \frac{\phi\sigma^2\sin(2d\alpha) + 2d\sinh(\sigma^2\alpha\phi)}{2 e^{3\alpha}[\cos(2d\alpha) + \cosh(\sigma^2\alpha\phi)]}, \label{eq:quantum-planar-1}\\
    \ell_{P}\dot{\phi} &= \frac{-\alpha\sigma^2\sin(2d\alpha) + 2d\cos(2d\alpha) + 2d\cosh(\sigma^2\alpha \phi)}{2 e^{3\alpha}[\cos(2d\alpha) + \cosh(\sigma^2\alpha\phi)]}. \label{eq:quantum-planar-2}
\end{align}

\begin{figure}
    \centering
    \includegraphics[width=0.5\textwidth]{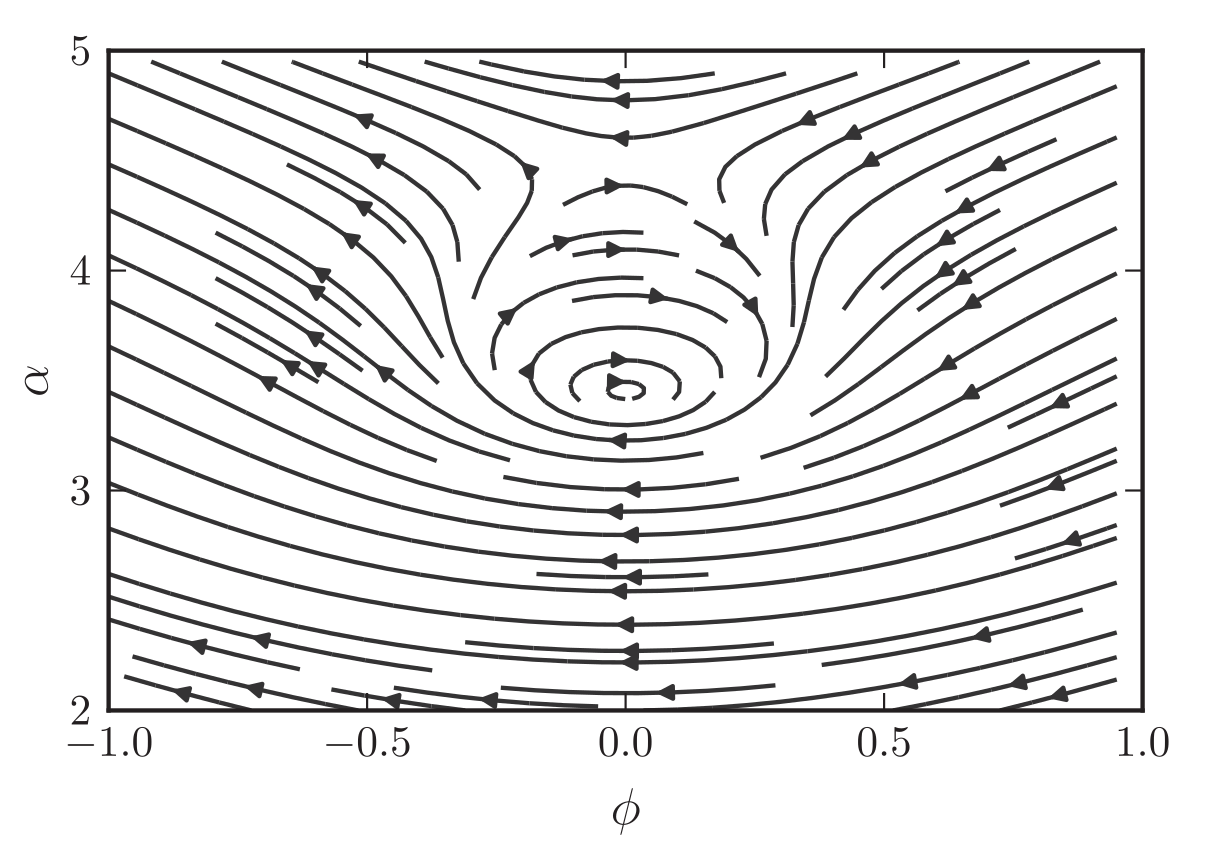}
    \caption{The dBB quantum trajectories for $d = -1$ and $\sigma = 1$. The bounce occurs when $\phi = 0$. Figure taken from reference \cite{Bacalhau:2017hja}.}
    \label{fig:quantum_phase_space}
\end{figure}

The numerical solution of this system leads to the Bohmian trajectories that we can see in the figure \ref{fig:quantum_phase_space}. There are bounce solutions that have a classical contraction and a classical expanding limit. These two phases will match with the classical solution we described in the previous subsection. There are also cyclic solutions that are non-physical because they do not have a classic limit.  

\subsection{Matching of the background}\label{sec:matching}

The matching of the classical and quantum background solutions when classical stiff matter evolution is attained ($w\approx 1$) in both cases was made with great details in \cite{Bacalhau:2017hja}. We just generalized the equations to include $\lambda$ as a free parameter of the model (see Appendix \ref{Appendix}), as in that work this parameter was fixed to $\lambda = \sqrt{3}$, imposing that the scalar field behaved as a perfect presureless matter fluid, $\omega = 0$, in the far past of the contracting phase. As a consequence of this choice, the model has an exact scale-invariant primordial power spectrum. Using the MCMC technique, we will let the data choose the value of this parameter, allowing the model to capture the near scale-invariant feature of the primordial power spectrum.   

The key free parameter, besides $\lambda$, that is important for the sequel of this work emerging from the analysis made in Ref.~\cite{Bacalhau:2017hja} is $\chi_b$, which gives the number of $e-$folds between the bounce and the moment when the Hubble radius is $1/(H_0 \sqrt{\Omega_d})$, where $H_0$ is the Hubble constant today, and $\Omega_d$ is the relative amount of dark matter in the contracting phase.

\section{Primordial perturbation}

We can derive the perturbed Einstein equations from the action of the model \cite{Falciano:2013uaa} 

\begin{equation}
    S = \int d\tau dx^3 \frac{z^2}{2}\left(\zeta^{\prime 2} + \frac{N^2 \zeta \Delta\zeta}{a^2}\right), \quad z^2 \equiv \frac{3 a^3 x^2}{\kappa^2 N},
    \label{eq:action-s}
\end{equation}
where $\zeta$ is the dimensionless gauge-invariant curvature perturbation, $\Delta$ is the spacial Laplacian, and the time derivative is defined as $^{\prime} \equiv d/d \tau$. The background quantities refers to the Bohmian quantum trajectories. Using the variational principle and the field decomposition, we can derive the following equation of motion for $\zeta$: 

\begin{equation}
    \zeta^{\prime \prime}_{k} + 2 \frac{z^{\prime}}{z}\zeta^{\prime}_{k} + \frac{N^2k^2}{R_{H}^{2} a^2}\zeta_{k} = 0,
\end{equation}
where $k$ is measured in units of $R_{H}^{-1}$, and the dimensionless time $\tau$ is given by  
\begin{equation}
    N = \frac{dt}{d\tau} = \frac{\tau}{H}.
\end{equation}

For the tensor perturbation, the action is similar to that of scalar perturbations 

\begin{equation}
    S = \int d\tau dx^3 \frac{z^2_{h}}{2}\left(h^{\prime 2} + \frac{N^2 h \Delta h}{a^2}\right), \quad z^2_{h} \equiv \frac{a^3 }{4 \kappa^2 N},
\end{equation}
with the equation of motion given by
\begin{equation}
    h^{\prime \prime}_{k} + 2 \frac{z^{\prime}_{h}}{z_{h}}h^{\prime}_{k} + \frac{N^2k^2}{R_{H}^{2} a^2}h_{k} = 0.
\end{equation}

The primordial power spectra of the scalar and tensor perturbations are

\begin{align}
    \Delta_{\zeta_{k}} &\equiv \frac{k^3|\zeta_{k}|^2}{R_{H}^{3} 2 \pi^2} = \frac{\ell_{P}^2}{R_{H}^{2}}\frac{4k^3 |\tilde{\zeta}_{k}|^2}{3 \pi}, \\
    \Delta_{h_{k}} &\equiv \frac{k^3|h_{k}|^2}{R_{H}^{3} 2 \pi^2} = \frac{\ell_{P}^2}{R_{H}^{2}}\frac{16k^3 |\tilde{h}_{k}|^2}{\pi},
\end{align}
where we define the dimensionless mode functions
\begin{equation}
    \zeta_{k} \equiv \sqrt{\frac{\kappa^2 R_{H}}{3}\tilde{\zeta_{k}}}, \quad h_{k} \equiv \sqrt{4\kappa^2 R_{H}}\tilde{h}_{k}.
    \label{eq:dimensionless-variables}
\end{equation}

The spectral indices for scalar and tensor perturbations (labels S and T) are given by
\begin{equation}
    n_{S,T} - 1 \equiv \left. \frac{d \log(\Delta_{\zeta_{k}, h_{k}})}{d \log k}  \right|_{k = k_{\ast}},
\end{equation}
and the tensor-to-scalar ratio is
\begin{equation}
    r \equiv \left.2\frac{\Delta_{h_{k}}}{\Delta_{\zeta}}\right|_{k = k_{\ast}}.
\end{equation}
The pivotal scale is set to $k_{\star} = 0.05R_{H}\text{Mpc}^{-1}$ as in the Planck collaboration paper \cite{ade2016planck}.

\subsection{Initial vacuum perturbations}

Using the Mukhanov-Sasaki variable $v \equiv z\zeta$, the equation of motion for scalar perturbation can be written as

\begin{equation}
    v^{\prime \prime}_{k} + \omega^2_{k}v_{k} = 0,
\end{equation}
where
\begin{equation}
    \omega_{k}^{2}(\eta,k) \equiv \frac{N^2 k^2}{a^2 R_{H}^2} - \frac{z^{\prime \prime}}{z}.
\end{equation}

We can find the following solution of the above equation through the WKB approximation formalism 

\begin{equation}
    \tilde{v}^{\text{WKB}}_{k} \approx \frac{1}{\sqrt{2 \omega_{k}}} e^{\pm i \int d\tau \omega_{k}}.
\end{equation}

In the far past, the modes are inside the horizon, $N^2k^2/(a^2R_{H}^{2}) \gg |z^{\prime \prime}/z|$, and the initial vacuum conditions are reduced to  
\begin{align}
    v_{\text{ini}} &= \frac{1}{\sqrt{2k}}\sqrt{\frac{a R_{H}}{N}},\\
    \left.\frac{dv}{d \tau}\right|_{\text{ini}} &= i \sqrt{2k}\sqrt{\frac{N}{a R_{H}}}.
\end{align}

The same procedure can be used for the tensor modes, with the Mukhanov-Sasaki variable defined as $\mu \equiv z_{h}h$. They satisfy equations similar to $v$ for the initial conditions. 

The numerical solutions for scalar and tensor perturbation are in a high oscillatory regime, which is very hard to integrate using the set of equations presented in this section. A better way to solve the equation of motion is to use action angles variables (AA) \cite{Celani:2016cwm}.

General linear oscillatory systems have a quadratic Hamiltonian in the form

\begin{equation}
    \mathcal{H} = \frac{\Pi^2_{\tilde{\zeta}_{k}}}{2m} + \frac{m \nu^{2}}{2}\tilde{\zeta}^{2}_{k},
\end{equation}
where $m$  is the associated ``mass'' of the system and $\nu$ is the frequency. The associated momenta of $\tilde{\zeta}_{k}$ are $\tilde{\Pi}_{k}$.

We can deduce the following expressions for $m$ and $\nu$ from the action of scalar perturbations \eqref{eq:action-s} and the dimensionless variable relations \eqref{eq:dimensionless-variables}:
\begin{equation}
    m = \frac{\kappa^{2}R_{H}z^{2}}{3} = \frac{a^3x^2 R_{H}}{N}, \quad \nu = \frac{Nk}{a R_{H}}.
\end{equation}

The complete set of equations in the action angle variables, and the relations needed to recover the original quantities
$\zeta$ and $h$ can be find in the Appendix of \cite{Bacalhau:2017hja}.

\subsection{Matter bounce model configurations}

In reference \cite{Bacalhau:2017hja} four parameter sets were presented that generate a scale-invariant primordial power spectrum with amplitudes of the order of $10^{-10}$. These sets are summarized in Table \ref{table:sets}. The parameters $d,\sigma,\alpha _b$ emerge from the wave function of the universe and an initial condition for the Bohmian trajectory. They are quantum parameters. The quantity $\chi_b$, as said above, is a free parameter emerging from the matching conditions. 

\begin{figure}[htbp]
    \centering
    \includegraphics[width=0.5\textwidth]{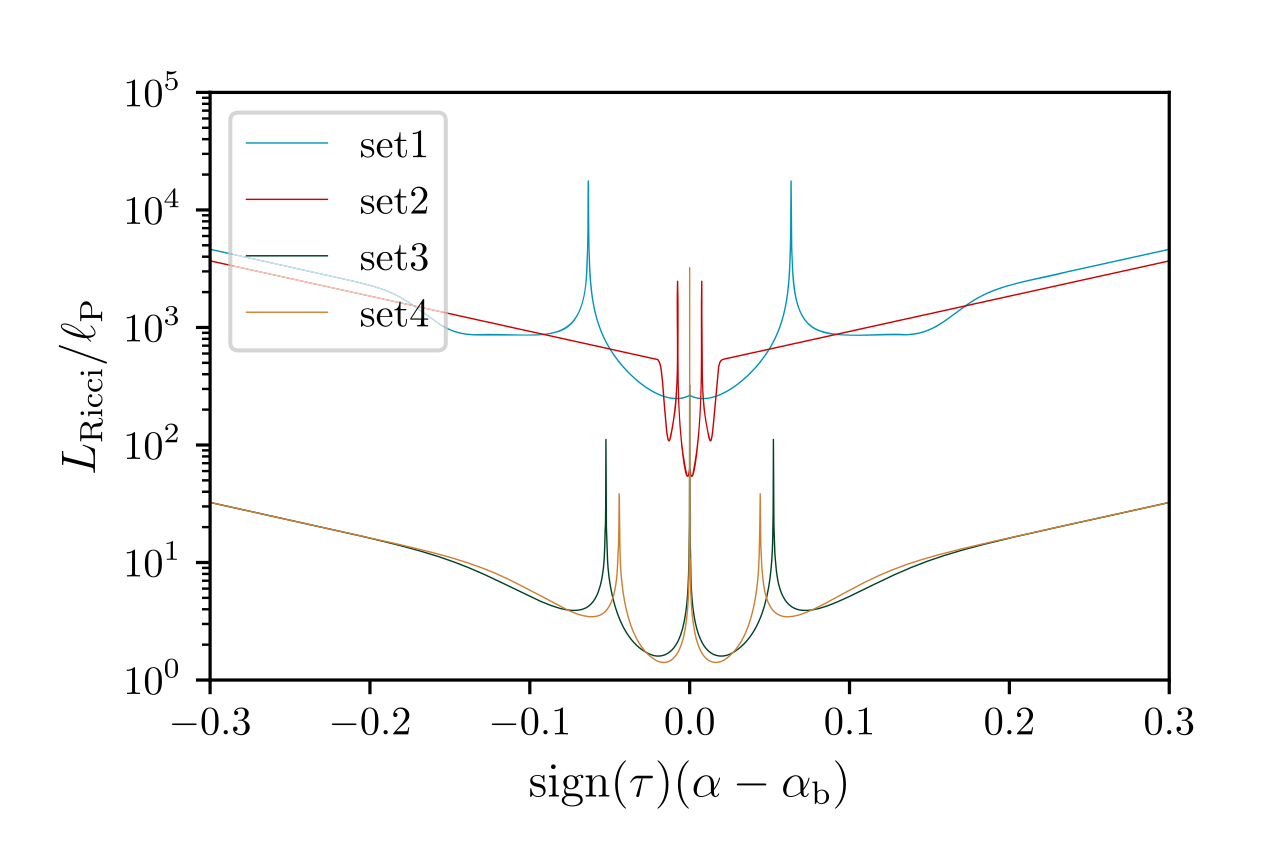}
    \caption{Time evolution of the Ricci scale for all sets appearing in Table \ref{table:sets}. The parameter d controls how close the scale gets to the Planck length, and set 3 and set 4 are in the limit of validity of our model. Figure taken from reference \cite{Bacalhau:2017hja}.}
    \label{fig:LR_LP}
\end{figure}

Note that only configurations 1 and 2 produce results for the Ricci scalar consistent with the constraints of a model based on the canonical quantization of gravity. As illustrated in Figure \ref{fig:LR_LP}, these two configurations yield an evolution for the Ricci scalar with values some orders of magnitude above the Planck scale; conversely, configurations 3 and 4 produce values approaching the Planck scale, which challenges our approximate approach to quantum gravity based on the Wheeler-DeWitt equation, requiring a more involved theory of quantum gravity, such as string theory and loop quantum gravity, which is out of the scope of this investigation.

\begin{table}[htbp]

\caption{Model parameters for four different cases in which the present model produces $\Delta \zeta$ close to $10^{-10}$ and scale-invariant spectra.}

\vspace{0.5cm}
{\normalsize
\begin{tabular}{lcccc}\toprule
\addlinespace[0.3ex]
\hline
\addlinespace[0.5ex]
           & d & $\sigma$ & $\alpha_{b}$ & $\mathcal{X}_{b}$\\

\addlinespace[0.5ex]

\hline

\addlinespace[0.5ex]

Set 1 & $-9\times 10^{-4}$ & 9   & $8.3163\times 10^{-2}$ & $2\times 10^{36}$   \\
Set 2 & $-9\times 10^{-4}$ & 100 & $7.4847\times 10^{-3}$ & $4\times 10^{36}$   \\
Set 3 & -0.1               & 4   & $10^{-5}$              & $6\times 10^{37}$   \\
Set 4 & -0.1               & 4   & $10^{-7}$              & $6\times 10^{37}$   \\

\addlinespace[0.5ex]

\bottomrule

\end{tabular}
}
\label{table:sets}
\end{table} 

Furthermore, configurations 1 and 2 generate a suppressed scalar-to-tensor ratio, $r = 1.9 \times 10^{-7} $ and $r = 1.3 \times 10^{-5} $, respectively, and these results are consistent with Planck 2018 observational data \cite{Planck:2018jri}. In this framework, the amplification of scalar perturbations is driven by quantum effects during the bounce, which is a novel feature of such models presented in Ref.~\cite{Bacalhau:2017hja}. In contrast, sets 3 and 4 fall outside the observational range due to their excessively high scalar-to-tensor ratios. Since sets 1 and 2 align with observational constraints regarding both the primordial power spectrum amplitude and the scalar-to-tensor ratio, they will serve as the basis for the Cosmic Microwave Background (CMB) angular power spectrum in our subsequent MCMC analysis using Planck 2018 data.

\section{Dataset and analysis methodology}

In this section, we present the data set and the methodology used to set constraints in the cosmological parameters of the inflationary $\Lambda$CDM model and the bounce models (set 1 and set 2).  

Our data sets include Planck 2018 CMB temperature, polarization, and lensing power spectra, with high-$\ell$ and low-$\ell$ TT, TE, EE and lensing likelihoods \cite{Planck:2019nip}.

In order to obtain the CMB power spectrum for the inflationary $\Lambda$CDM and the bounce models, we use the public code CLASS \cite{Diego_Blas_2011}. The primordial power spectra of the bounce models were calculated using the NcHICosmoVexp code, that is part of NumCosmo (numerical cosmology library) \cite{Vitenti2012c}. It was not necessary to modify CLASS to include the bounce models, because it has a prescription to call external primordial power spectra when they are different from the standard inflationary model. The NcHICosmoVexp code was modified using the equations, coming from the matching conditions described in Ref.~\cite{Bacalhau:2017hja}, now allowing different values for the $\lambda$ parameter.

To set constraints on the cosmological parameters, the models were compared with the Planck 2018 data through the Markov Chain Monte Carlo (MCMC) method, using the Metropolis-Hastings algorithm implemented in the Cobaya code \cite{Torrado:2020dgo}. We use GetDist \cite{Lewis:2019xzd} to make statistical analysis of the MCMC samples and to plot the parameters $1\sigma$ and $2\sigma$ confidence contours. 

For the inflationary $\Lambda$CDM model, our MCMC analysis utilizes the standard parameter set $\{ H_0,\Omega_{b} h^{2}, \Omega_{c} h^{2}, \tau, A_{s}, n_s \}$. In contrast, the bounce model configurations (set 1 and set 2) are sampled using $\{H_{0},\Omega_{b} h^{2}, \Omega_{c} h^{2}, \tau, \mathcal{X}_{b},\lambda \}$. The prior distributions for these parameters are summarized in Table \ref{table:prior}. For the parameters $\mathcal{X}_{b}$ and $\lambda$ we choose a numerical stable range  around the fiducial values of set 1 and set 2 for these parameters. Notably, we have elected to fix the wavefunction parameters $d$ and $\sigma$ — representing the average scale and dispersion of the universe's wavefunction — at their respective fiducial values. In the bounce framework, $d$ and $\sigma$ are coupled with the parameter $\mathcal{X}_{b}$ to determine the primordial power spectrum amplitude, while $\lambda$ dictates the spectral index. By fixing the wavefunction parameters, we effectively reduce the dimensionality of the parameter space, ensuring that all models are compared using an equal number of free parameters, providing a consistent statistical baseline.

\begin{table}[htbp]
\caption{The priors for each cosmological parameter. The bounce model, set 1 and 2, use the same priors for coincident parameters of the $\Lambda$CDM model, same prior for the parameter $\lambda$, and different prior for the $\mathcal{X}_{b}$ parameter.}

{\normalsize
\begin{tabular}{ccc}\toprule
\addlinespace[0.3ex]
\hline
\addlinespace[0.5ex]

Model              &Parameter           & Priors                                       \\ 
\midrule                              
Common parameters  & $\Omega_{b}h^{2}$   & $[0.005,0.1]$ \\
                   & $\Omega_{c}h^{2}$   & $[0.001,0.99]$ \\
                   & $H_{0}$             & $[50,90]$ \\
                   & $\tau$              & $[0.01,0.8]$ \\
\midrule
$\Lambda$CDM       & $\ln(10^{10}A_{s})$ & $[1.61, 3.91]$ \\
                   & $n_{s}$             & $[0.8, 1.2]$ \\
\midrule
Set 1              & $\mathcal{X}_{b}$   & $[1.0, 2.5]$ \\
                   & $\lambda$           & $[1.72, 1.734]$ \\
\midrule
Set 2              & $\mathcal{X}_{b}$   & $[3.5, 4.5]$ \\
                   & $\lambda$           & $[1.72, 1.734]$ \\
\bottomrule
\end{tabular}
}
\label{table:prior}
\end{table} 

\section{Results}

\begin{figure*}[htbp]
    \centering
    \includegraphics[width=1.0\textwidth]{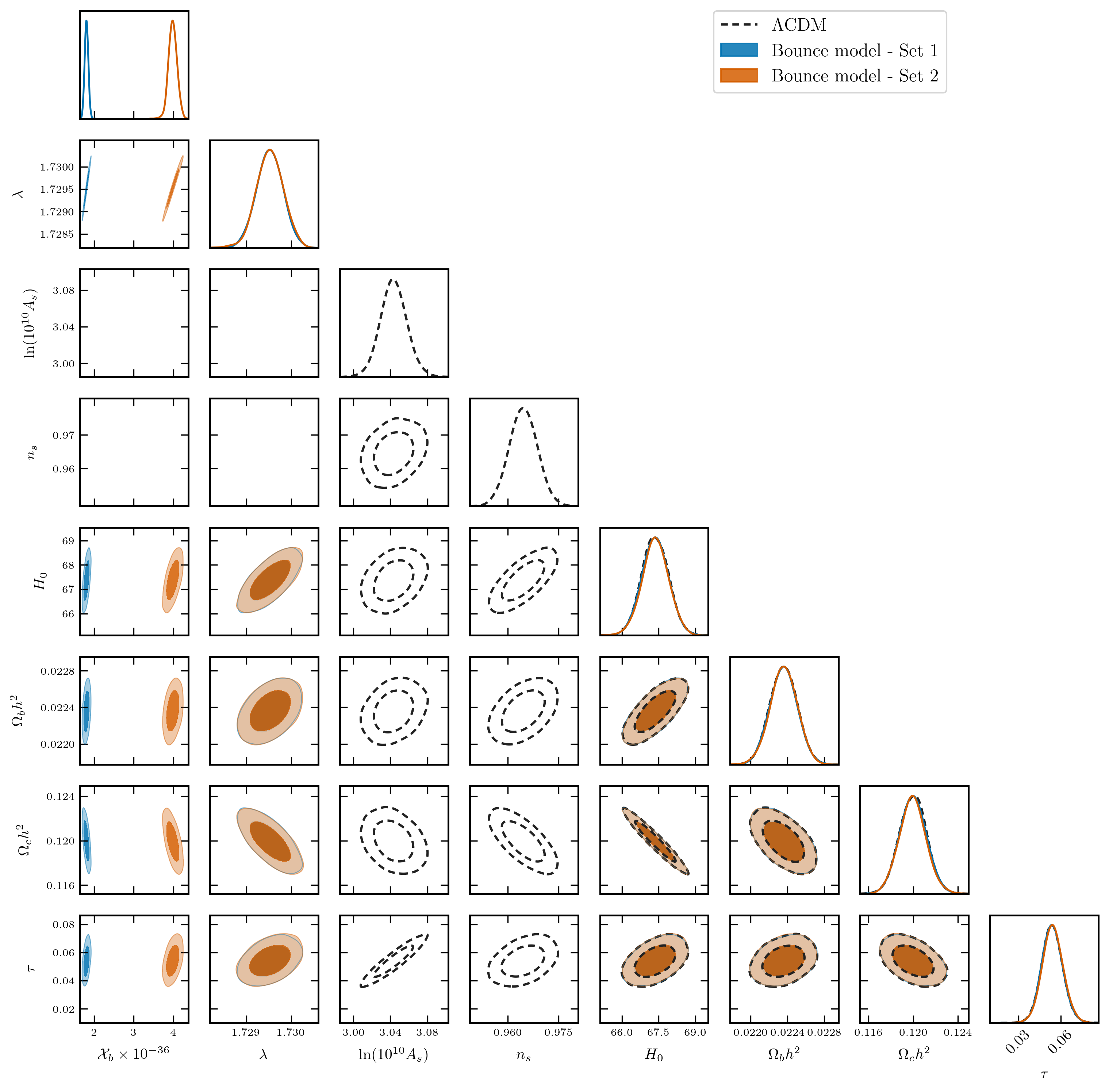}
    \caption{Constraints on parameters of the $\Lambda$CDM model and bounce models, set 1 and set 2, from the {\it Planck 2018 EE, TE} and {\it TT} high-$\ell$ spectra combined with {\it Planck 2018 TT} and {\it EE} low-$\ell$, and, with {\it Planck 2018 lensing}. Contours contains 68$\%$ and 95$\%$ of the probability.}
    \label{fig:triangle_plot}
\end{figure*}

Utilizing a Markov Chain Monte Carlo (MCMC) analysis constrained by Planck 2018 data, we derived posterior distributions for the cosmological parameters across three scenarios: the standard inflationary model and two distinct bounce configurations (set 1 and set 2). The resulting parameter constraints are summarized in Table \ref{table:constraints}, with the corresponding marginalized likelihoods and correlations illustrated in the triangle plot of Figure \ref{fig:triangle_plot}.

In Figure \ref{fig:triangle_plot}, the solid and shaded regions represent the 1$\sigma$ and 2$\sigma$ confidence levels, respectively. Inflationary $\Lambda$CDM, bounce set 1, and bounce set 2 models are distinguished by dashed contours, blue, and orange regions. The three models yield highly consistent posterior distributions for the primary parameters $H_{0}$, $\Omega_{b}h^2$, $\Omega_{c}h^2$, and $\tau$. Furthermore, both bounce sets provide nearly identical constraints for the parameter $\lambda$. 

\begin{table*}[htbp]

\centering
\caption{Parameter 68 $\%$ intervals and the minimum $\chi^2$ for the $\Lambda$CDM model, and matter bounce model - ser 1 and set 2 - from the {\it Planck 2018 EE, TE} and {\it TT} high-$\ell$ spectra combined with {\it Planck 2018 TT} and {\it EE} low-$\ell$, and, with {\it Planck 2018 lensing}. }

\vspace{0.5cm}

{\normalsize
\begin{tabular}{llll}\toprule
\addlinespace[0.3ex]
\hline
\addlinespace[0.5ex]
Parameter & $\Lambda$CDM & Set 1 & Set 2\\

\addlinespace[0.5ex]

\hline

\addlinespace[0.5ex]

$\Omega_{b}h^{2}$                   &0.02236 $\pm$ 0.00015 & 0.02236 $\pm$ 0.00015   & 0.02236 $\pm$ 0.00015   \\
$\Omega_{c}h^{2}$                   &0.1200  $\pm$ 0.0012  & 0.1199  $\pm$ 0.0012    & 0.1199  $\pm$ 0.0012    \\
$H_{0}$                             &67.36 $\pm$ 0.57      & 67.37 $\pm$ 0.55        & 67.38 $\pm$ 0.56        \\
$\tau$                              &0.0540 $\pm$ 0.0075   & 0.0542 $\pm$ 0.0078     & 0.0543 $\pm$ 0.0076     \\
$\ln(10^{10}A_{s})$                 &3.044 $\pm$ 0.015     & \multicolumn{1}{c}{-}   & \multicolumn{1}{c}{-}  \\
$n_{s}$                             &0.9646 $\pm$ 0.0042   & \multicolumn{1}{c}{-}   & \multicolumn{1}{c}{-}  \\
$\mathcal{X}_{b} \times 10^{-36}$   & \multicolumn{1}{c}{-}& 1.808 $\pm$ 0.044       & 3.99 $\pm$ 0.10       \\ 
$\lambda$                           & \multicolumn{1}{c}{-}& 1.72953 $\pm$ 0.00028   & 1.72953 $\pm$ 0.00029   \\ 

\addlinespace[0.5ex]
\hline
\addlinespace[0.5ex]

$\chi^2$                           & 2773.30               & 2772.41                 & 2772.99                \\ 

\bottomrule

\end{tabular}
}
\label{table:constraints}
\end{table*} 

\begin{figure}[htbp]
	\centering
	\begin{subfigure}{1.0\linewidth}
		\includegraphics[width=\linewidth]{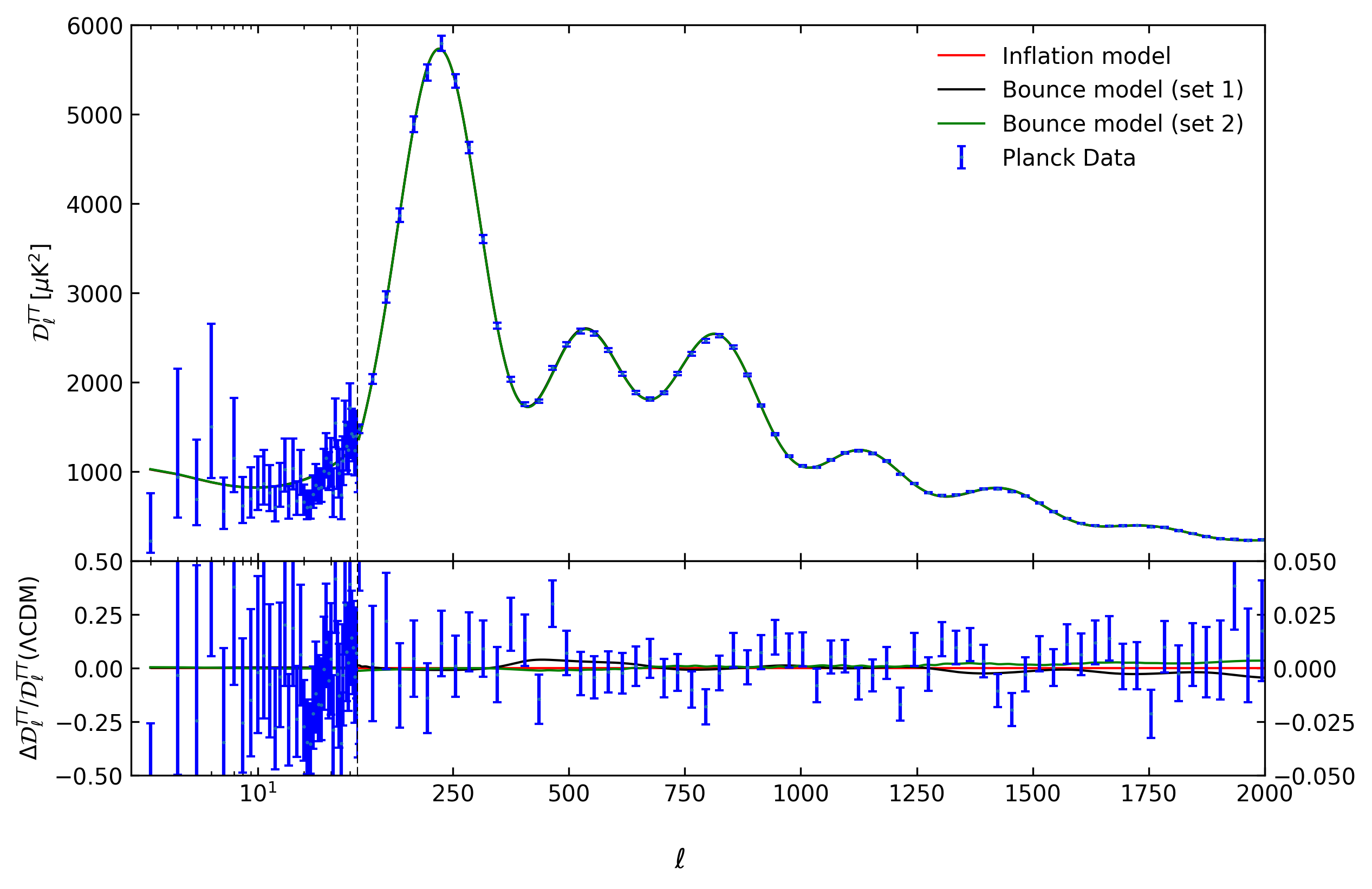}
		\label{fig:TT}
	\end{subfigure}
    \begin{subfigure}{1.0\linewidth}
		\includegraphics[width=\linewidth]{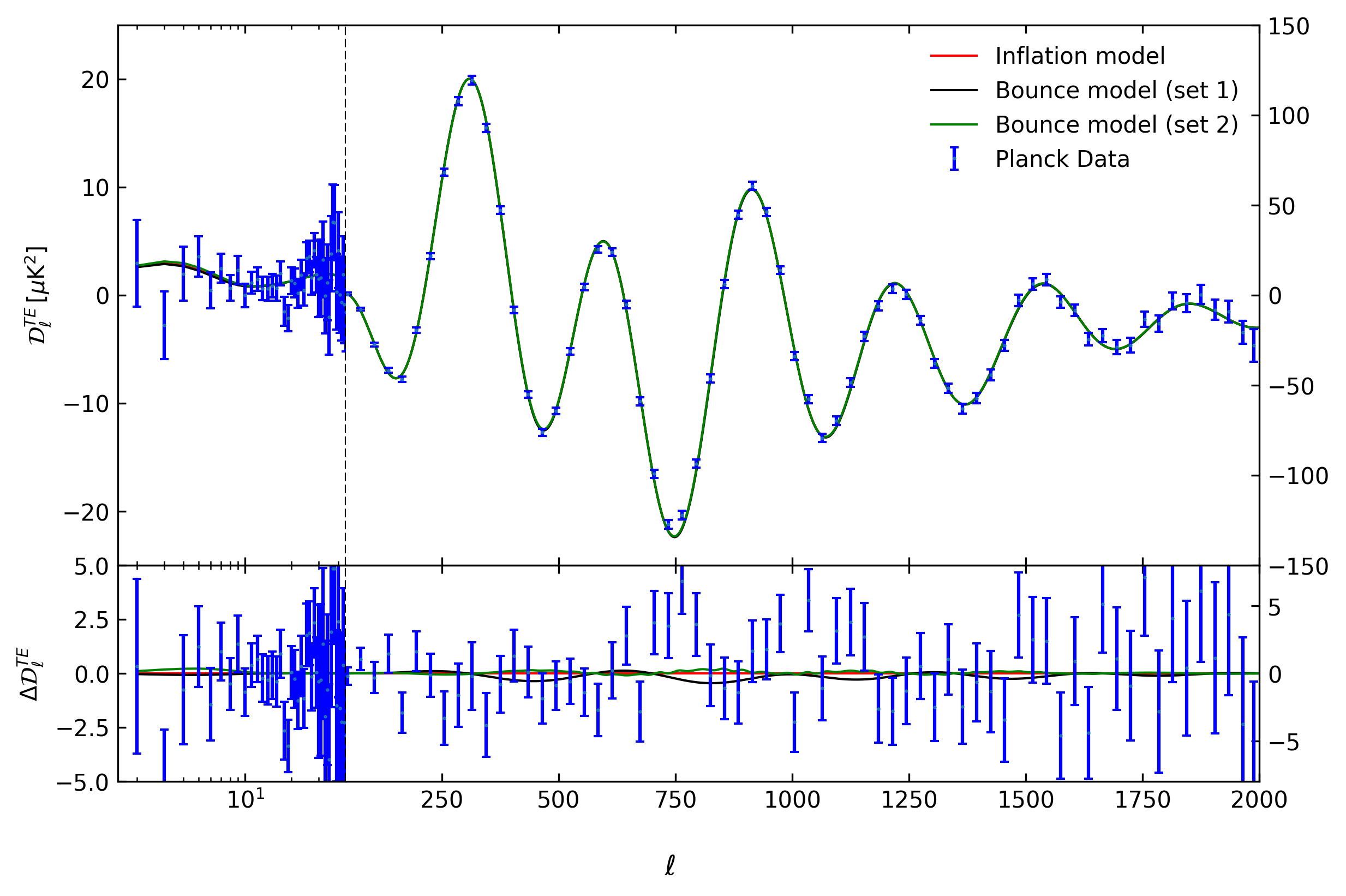}
		\label{fig:TE}
	\end{subfigure}
	\begin{subfigure}{1.0\linewidth}
		\includegraphics[width=\linewidth]{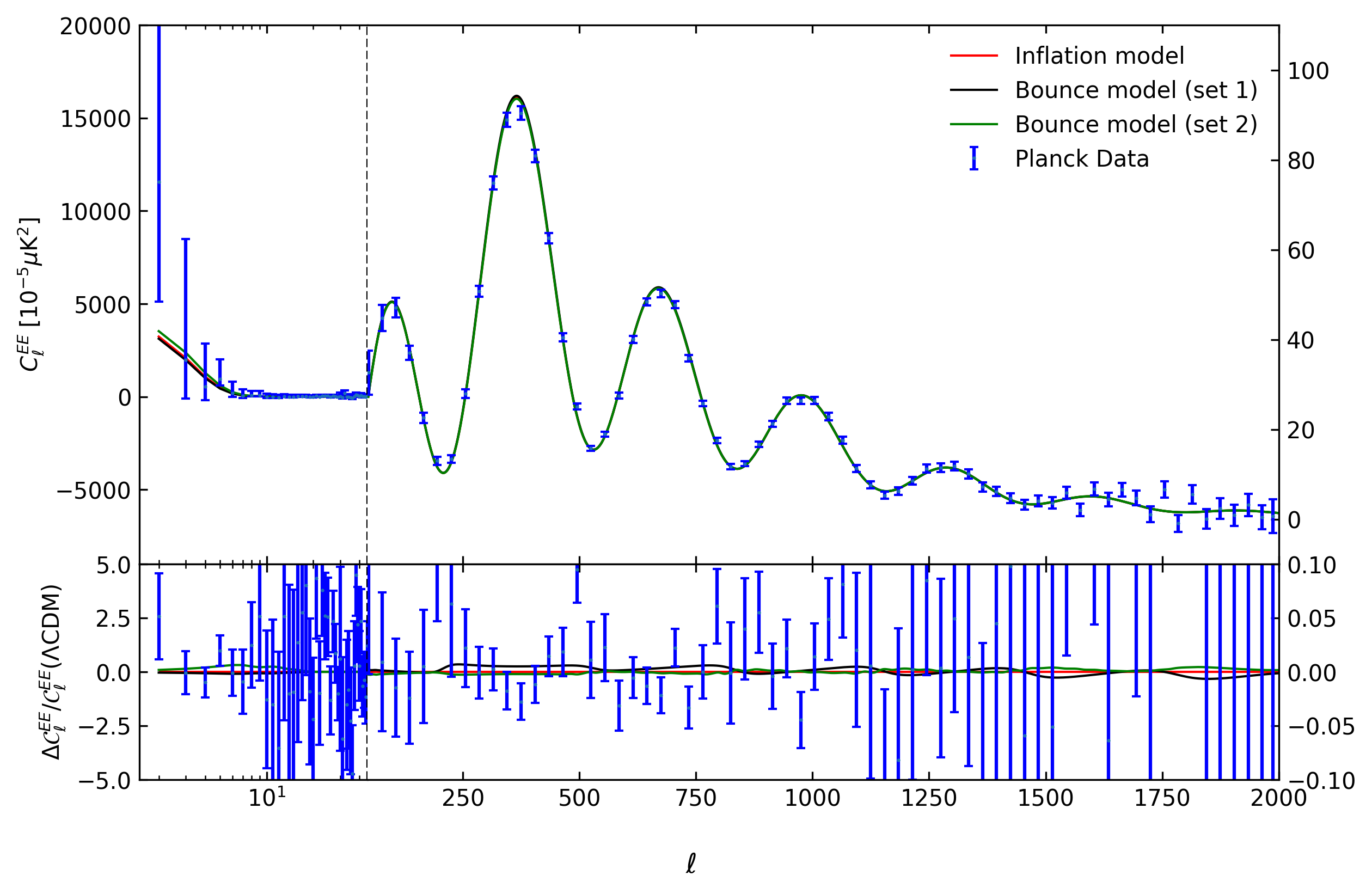}
		\label{fig:EE}
	\end{subfigure}
	     
	\caption{For the three plots, the top panels shows the TT, TE and EE power spectra, respectively, of the inflationary $\Lambda$CDM model and bounce models, set 1 and set 2, using the best-fit values from the MCMC analyses. In the bottom panels are the TT, TE and EE residuals power spectra of the models with respect to the inflationary $\Lambda$CDM model.}
	\label{fig:TTTEEE}
\end{figure}

The best-fit parameters were obtained by minimizing the negative log-likelihood using the Cobaya minimization package. Since flat priors were adopted for all sampled parameters, this procedure yields the Maximum Likelihood Estimation (MLE) \cite{Trotta:2017wnx}. With these best-fit values, we computed the theoretical TT, TE, and EE angular power spectra for the three models, as shown in Figure \ref{fig:TTTEEE}. The inflationary $\Lambda$CDM model and the two bounce configurations (set 1 and set 2) are represented by red, black, and green curves, respectively, overlaid with the Planck 2018 data and associated 1$\sigma$ uncertainties.

As shown in the figure, both bounce configurations provide a fit to the data that is statistically indistinguishable from the standard inflationary $\Lambda$CDM model. The lower subpanels display the residuals of the bounce models relative to the $\Lambda$CDM best-fit. While the residuals highlight minor deviations in the predicted power spectra across the TT, TE, and EE channels, these differences remain well within the experimental error bars. 

To evaluate the relative fit of the inflationary $\Lambda$CDM model compared to the two configurations of the matter bounce model (set 1 and set 2) against the Planck 2018 data, we employ the Akaike Information Criterion (AIC) and the Bayesian Information Criterion (BIC) \cite{Liddle:2007fy}. These are defined as $\text{AIC} = -2 \ln{\mathcal{L}_{\text{max}}} + 2k $ and $\text{BIC} = -2 \ln{\mathcal{L}_{\text{max}}} + k \ln{N}$, where $\mathcal{L}_{\text{max}}$ represents the maximum likelihood, $k$ is the number of free parameters, and $N$ denotes the number of data points. Given that the models under consideration possess an identical number of parameters, the information criteria analysis simplifies to a comparison of the minimum chi-square ($\chi^{2}_{\text{min}}$), assuming a Gaussian likelihood distribution where $-2 \ln{\mathcal{L}_{\text{max}}} \approx \chi^{2}_{\text{min}}$.

The values for $\chi^{2}_{\text{min}}$ across the three models are presented in the last line of Table \ref{table:constraints}. Taking the inflationary model as the reference, we find that the difference in chi-square is $\Delta \chi^2 = -0.89$ for set 1 and $\Delta \chi^{2} = -0.31$ for set 2. While these negative values indicate a marginally improved fit for the matter bounce model configurations, the models remain statistically indistinguishable, as the magnitude of $|\Delta\chi^2| < 2$ in both cases. This suggests that current Planck data does not possess sufficient sensitivity to discriminate between these early-universe scenarios based solely on two-point statistics.

\begin{figure}[htbp]
    \centering
    \includegraphics[width=0.5\textwidth]{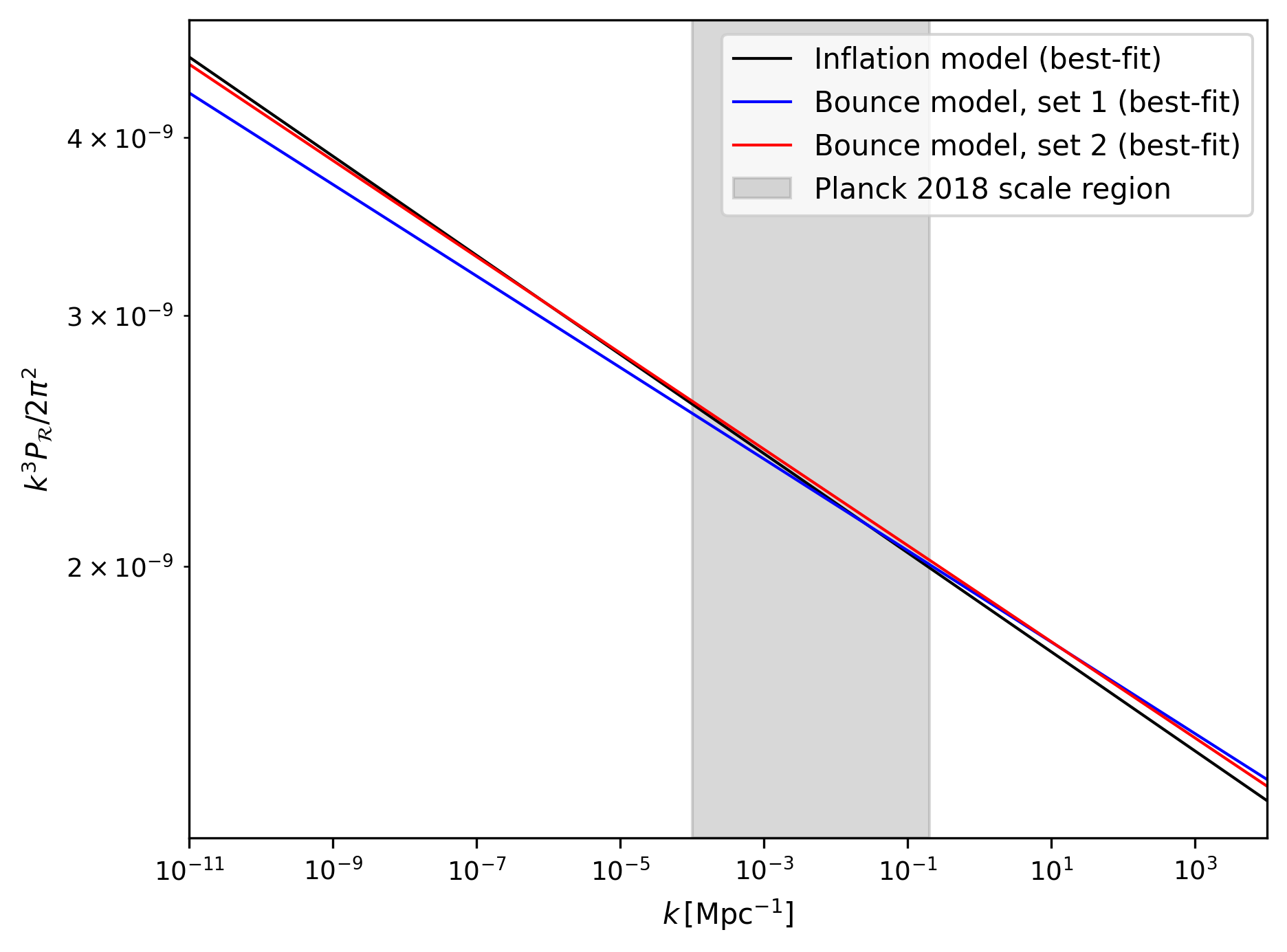}
    \caption{Primordial power spectrum of the inflationary $\Lambda$CDM model and bounce models, set 1 and set 2, in the best-fit values from the MCMC analyses. The gray band shows the region observed by the Planck satellite. }
    \label{fig:Pk}
\end{figure}

To investigate the underlying phenomenology of these models, we compare their primordial power spectra, $P(k)$, using the maximum likelihood estimates derived from our analyses. In Figure \ref{fig:Pk}, the spectra for the inflationary $\Lambda$CDM model and the two bounce configurations (set 1 and set 2) are represented by black, blue, and red curves, respectively. The shaded gray region identifies the range of co-moving scales (k) effectively probed by the Planck satellite.

\section{Conclusion}

We have seen that the primordial linear perturbation initial conditions emerging from the viable scalar field matter bounce scenarios proposed in Ref.~\cite{Bacalhau:2017hja} yields CMB anisotropies that fit very well with Planck 2018 data, and that their predictions are statistically indistinguishable from the predictions coming from inflation.

In this matter bounce scenario — characterized by a single-fluid evolution and a subsequent dark energy era — the spectral index $n_s$ is primarily governed by the parameter $\lambda$. As illustrated, all three models predict nearly parallel power-law spectra with $n_s \approx 0.965$ within the Planck observational window. However, the amplitude of the bounce spectrum is a derived quantity dependent on a broader parameter space, including the Hubble constant $H_{0}$, the dark energy density $\Omega_{\Lambda}$, and the $\mathcal{X}_{b}$ parameter (with the wavefunction parameters $d$ and $\sigma$ held fixed in this study).

This multi-parameter dependence grants the bounce model significant flexibility in matching the observed CMB normalization. While the primordial power spectra converge within the scales constrained by Planck, they diverge significantly when extrapolated to the infrared (low k) and ultraviolet (high k) regimes. Consequently, distinguishing between these competing scenarios will require data from high-resolution experiments — such as the Atacama Cosmology Telescope (ACT) \cite{AtacamaCosmologyTelescope:2025blo}, the South Pole Telescope (SPT) \cite{SPT:2004qip}, and the forthcoming CMB-S4 mission \cite{Abazajian:2019eic} — which probe the smaller comoving scales (high k) beyond the sensitivity of Planck. Furthermore, an analysis of higher-order statistics, such as the bispectrum, will be essential to break the functional degeneracies inherent in the two-point power spectrum and isolate the unique non-Gaussian signatures of the matter bounce.

Furthermore, such models predict that the spectrum of gravitational waves deviate from the almost scale invariant spectrum at high frequencies ($k^3$ or even $k^4$), which may increase significantly its power at high frequencies, opening the possibility of being detected in future experiments able to measure stochastic gravitational waves, as LISA \cite{LISA:2017pwj, LISACosmologyWorkingGroup:2022jok} and LIGO Advanced \cite{LIGOScientific:2014pky}. These are also viable possibilities to distinguish such models from inflation. 

\begin{acknowledgments}

RFP and NPN acknowledge the support of CNPq of Brazil under grants 140696/2022-9 and PQ-IB 310121/2021-3, respectively.
\end{acknowledgments}

\appendix

\section{Matching of background: generalization}\label{Appendix}

The match of classical and quantum background solutions was made with great details in \cite{Bacalhau:2017hja}. Here we will generalize the equations to include $\lambda$ as a free parameter of the model. In that work, this parameter was fixed to $\lambda = \sqrt{3}$, allowing the scalar field to behave as a matter fluid, $\omega = 0$, in the far past of the universe contracting phase. As a consequence of this choice, the model has an exact scale-invariant primordial power spectrum. Using the MCMC technique, we let the data choose the value of this parameter, allowing the model to capture the near scale-invariant feature of the primordial power spectrum.   

The classical contracting solution starts when $x \approx \lambda/\sqrt{6}$ and ends in $x \rightarrow 1$. The quantum contracting phase starts in $x\approx 1$ and bounces to $x \rightarrow -1$, in the expanding phase. Finally, the  classical expanding solution starts in $x \approx -1$ and ends with $x \rightarrow \lambda/\sqrt{6}$. In order to properly match the classical and quantum solutions, we must parameterize $x$ in the vicinity of the critical points, $S_{-}$ and $S_{+}$, by $x = \pm(1-\epsilon_{\pm})$, $0 < \epsilon_{\pm} \ll 1$. The quantum solution has classical limits, which allows for a continuous connection of these solutions.

We will rewrite the contraction and expansion matching parameters, $\epsilon_{c}$ and $\epsilon_{e}$, respectively, in terms of new parameters that controls the number of e-folds between the bounce and the Hubble scale. In order to do this, we solved the planar system, equations \eqref{eq:planar-1} and \eqref{eq:planar-2}, and got implicit functions for $x(\alpha)$ and $H(\alpha)$:

\begin{equation}
\label{eq:implicit-1}
3\alpha = - \frac{3}{6-\lambda^{2}}\ln{\left[\frac{\left(\frac{\lambda}{\sqrt{6}} - x\right)^2}{1-x^2}\right]} - \frac{\sqrt{6}\lambda\tanh^{-1}{(x)}}{6- \lambda^2} + \text{const},
\end{equation}

\begin{equation}
\label{eq:implicit-2}
    \ln{H} = \frac{\lambda^2}{6 - \lambda^2}\ln{\left[\frac{\left(\frac{\lambda}{\sqrt{6}} - x\right)}{(1 - x^2)^\frac{3}{\lambda^2}}\right]} + \frac{\sqrt{6}\lambda}{6-\lambda^2}\tanh^{-1}(x) + \text{const}.
\end{equation}

The number of e-folds between the bounce and the Hubble scale is better determined by writing these equations in the following way:

\begin{align}
    \left(\frac{a}{\bar{a}_{0}}\right)^6\left(\frac{H}{H_{0}}\right)^2 &= \frac{C_{1}}{\left(\frac{\lambda}{\sqrt{{6}} } - x\right)^2}, \label{eq:aH}\\
    \left(\frac{a}{\bar{a}_{0}}\right)^3 &= \frac{C_{2}(1-x)^{\gamma_{+}}(1+x)^{\gamma_{-}}}{\left(\frac{\lambda}{\sqrt{6}} - x\right)^{4\gamma_{+}\gamma_{-}}}, \label{eq:a}
\end{align}
where we defined $\gamma_{\pm} \equiv 1/\left[2\left(1 \mp \frac{\lambda}{\sqrt{6}}\right)\right]$ with a signal convention which recovers $\lambda_{\pm} = 1 \pm \frac{1}{\sqrt{2}}$, when $\lambda = \sqrt{3}$. $C_{1}$ and $C_{2}$ are constants, $H_{0}$ is the Hubble constant today, and $\bar{a}_{0}$ is a  scale factor constant that can be chosen conveniently. $H_{0}$ and $\bar{a}_{0}$ can be absorbed in the constants $C_{1}$ and $C_{2}$. Expanding the equations \eqref{eq:aH} and \eqref{eq:a} around the critical points $x = \pm (1 - \epsilon_{\pm})$ and $x = \lambda/\sqrt{6} \pm \epsilon_{\lambda}$ leads us to the equations

\begin{align}
    \left(\frac{a_{\pm}}{\bar{a}_{0}}\right)^3 &= C_{2}2^{\gamma_{\mp}}\epsilon_{\pm}^{\gamma_{\pm}} (2\gamma_{\pm})^{4\gamma_{+}\gamma_{-}}, \label{eq:a_pm_a0}\\
    \left(\frac{a_{\lambda}}{\bar{a}_{0}}\right)^3 &=\frac{C_{2}}{(2\gamma_{+})^{\gamma_{+}}(2\gamma_{-})^{\gamma_{-}}\epsilon_{\lambda}^{4\gamma_{+}\gamma_{-}}}. \label{eq:a_lambda_a_0}
\end{align}

We can relate the contraction matching parameter with the number of e-folds between the bounce and the Hubble scale when dividing these two equations.

\begin{equation}
    \left(\frac{a_{+}}{a_{\lambda}}\right)^3 = 2^{\gamma_{-}}(2\gamma_{+})^{4\gamma_{+}\gamma_{-}}(2\gamma_{+})^{\gamma_{+}}(2\gamma_{-})^{\gamma_{-}} \epsilon_{c}^{\gamma_{+}}\epsilon_{\lambda}^{4\gamma_{+}\gamma_{-}},
\end{equation}

In a similar way, we can calculate the number of e-folds to the dark energy phase, when $x=0$, yielding

\begin{align}
\left(\frac{a_{de}}{\bar{a}_{0}}\right)^3 &= \frac{C_{2}}{\left(\frac{\lambda}{\sqrt{6}}\right)^{4\lambda_{+}\lambda_{-}}}, \\
\left(\frac{a_{-}}
{a_{de}}\right)^3 &= 2^{\gamma_{+}}\epsilon_{e}^{\gamma_{-}}(2\gamma_{-})^{4\gamma_{+}\gamma_{-}}\left(\frac{\lambda}{\sqrt{6}}\right)^{4\lambda_{+}\lambda_{-}},   
\end{align}
where we replaced $\epsilon_{+}$ and $\epsilon_{-}$ by $\epsilon_{c}$ and $\epsilon_{e}$, respectively, because this is the meaning of them in the model with dark energy era in the expanding phase.

\subsection{Initial conditions at the bounce}

To construct the background solutions, we start by integrating the quantum planar system in the expanding phase and the contracting phase with initial conditions near the bounce. For each phase, we evolve the quantum system until the classical limits. The final state of the quantum system will be the initial condition for the classical system. To do this, we define a new time variable

\begin{equation}
\label{eq:alpha_tau}
    \alpha = \alpha_{b} + \frac{\tau^2}{2}.
\end{equation}
Setting the bounce at $\tau = 0$, the parameter $\alpha_{b}$ becomes the initial condition for $\alpha$ in this coordinate system.   

In the bounce we have the maximum contraction of the universe, when the Hubble factor is zero, $ \dot{\alpha} = 0$. Using this information in the first quantum planar system \eqref{eq:quantum-planar-1}, we can see that $\phi$ must be zero to cancel the right hand side of the equation.

Expanding the quantum planar system, \eqref{eq:quantum-planar-1} and \eqref{eq:quantum-planar-2}, around the bounce we have:

\begin{align}
    \frac{d\tau}{dt_{Q}} &= \frac{\phi}{\tau}D_{1} \label{eq:quantum-bounce-1}\\
    \frac{d\phi}{dt_{Q}} &= D_{2}, \label{eq:quantum-bounce-2} 
\end{align}
where $t_{Q}$ is a convenient dimensionless time variable, given by $e^{3 \alpha}\ell_{p}dt_{Q} = dt$. The constants $D_{1}$ and $D_{2}$ are
\begin{align}
    D_{1} &=\frac{\sigma^2[\sin{(2d\alpha_{b}) + 2d\alpha_{b}}]}{2[2\cos(2d\alpha_{b}) + 1]}, \\
    D_{2} &=\frac{-\alpha_{b}\sigma^2\sin{(2d\alpha_{b}) + 2d\cos{(2d\alpha_{b}) + 2d}}}{2[2\cos(2d\alpha_{b}) + 1]}.
\end{align}

The quantum system around the bounce, \eqref{eq:quantum-bounce-1} and \eqref{eq:quantum-bounce-2}, can be integrated and has the following parametric solution
\begin{equation}
    \tau = t_{Q}\sqrt{D_{1}D_{2}}, \quad \phi = t_{Q}D_{2}.
\end{equation}
The numerical integration begins close to the bounce, with the dimensionless time variable chosen to be very small, $t^{\text{ini}}_{Q} \propto \pm \mathcal{O}(10^{-50})$. From the above equation, we can substitute $\tau$ in \eqref{eq:alpha_tau}, hence obtaining the initial conditions for $\phi$ and $\alpha$.

\subsection{Matching with matter-domination scale}

The classical evolution begins at the end of the quantum evolution. We can match the solutions through the combination of the classical limit of the quantum planar system, \eqref{eq:quantum-planar-1} and \eqref{eq:quantum-planar-2}, that is 
\begin{align}
    x & \approx \coth(\sigma^2 \alpha \phi),\label{eq:classical-limit-1}\\
    \frac{H}{H_{0}} & \approx \frac{R_{H}}{\ell_{P}}\frac{d e^{-3\alpha}}{x}, \label{eq:classical-limit-2}\\
    \ell_{P}\dot{\phi} & \approx d e^{-3\alpha},\label{eq:classical-limit-3}
\end{align}
with the implicit functions for $x(\alpha)$ and $H(\alpha)$, equations \eqref{eq:implicit-1} and \eqref{eq:implicit-2}, respectively. 

In the contracting phase, the match happens close to the stiff-matter critical point, when $x(a_{+}) = 1-\epsilon_{+}(a_{+})$. Setting $\bar{a}_{0}$ as the scale factor in the far past, it must to satisfy the condition $x(\bar{a}_{0}) = \lambda/\sqrt{6} + \epsilon_{\lambda}(\bar{a}_{0})$. Now, we expand the equation \eqref{eq:implicit-1} about the matching point, and equate it with the equation \eqref{eq:classical-limit-2}, this yields the following expression for the constant $C_{1}$
\begin{equation}
    C_{1} = \left(\frac{R_{H}}{\ell_{P}}\right)^2 \frac{d^2}{(2\gamma_{+})^2 (\mathcal{X}_{b}a_{b})^6},
\end{equation}
where $\mathcal{X}_{b}$ is a reparameterization of $\bar{a}_{0}$ through the definition $\mathcal{X}_{b} \equiv \bar{a}_{0}/a_{b}$. This new parameter give to us the number of e-folds from the bounce to $\bar{a}_{0}$, the scale factor when the universe is contracting at a rate given by the negative of the Hubble constant today.

We obtain the constant $C_{2}$ from the equation \eqref{eq:a_pm_a0},
\begin{equation}
    C_{2} = \frac{1}{2^{\gamma_{-}}(2\gamma_{+})^{4\gamma_{+}\gamma_{-}}\epsilon_{+}^{\gamma_{+}}}\left(\frac{a_{+}}{\bar{a}_{0}}\right)^3.
    \label{eq:C2} 
\end{equation}
This equation relates the constant $C_{2}$ to $a_{+}$ and $\epsilon_{+}$. At the end of the quantum regime, $a_{+}$ can be any value satisfying $0 < \epsilon_{+}\ll 1$. Instead of choosing a specific value for $a_{+}$ and $\epsilon_{+}$, the constant $C_{2}$ can be rewritten in terms of cosmological parameters imposing a condition for the end of the quantum evolution.

Expanding the equations \eqref{eq:aH} and \eqref{eq:a} about the far past critical point $x_{\lambda} = \lambda/\sqrt{6} + \epsilon_{\lambda}$, yields

\begin{equation}
    \left(\frac{H_{\lambda}}{H_{0}}\right)^2 \approx C_{1}\left[\frac{(2\gamma_{+})^{\gamma_{+}}(2\gamma_{-})^{\gamma_{-}}}{C_{2}}\right]^{\frac{1}{2\gamma_+ \gamma_-}} \left(\frac{\bar{a}_{0}}{a_{\lambda}}\right)^{3\left(2 - \frac{1}{2\gamma_{+}2\gamma_{-}}\right)}.
\end{equation}
When $a_{\lambda} = \bar{a}_{0}$, the Hubble factor in the far past can be related with the Hubble constant today through the new constant $\Omega_{d}$ as $H^2(a_{\lambda} = \bar{a}_{0}) \approx \Omega_{d} H_{0}^2$, yielding, 

\begin{align}
    \nonumber \Omega_{d} &\equiv C_{1}\left[\frac{(2\gamma_{+})^{\gamma_{+}}(2\gamma_{-})^{\gamma_{-}}}{C_{2}}\right]^{\frac{1}{2\gamma_+ \gamma_-}} \\ 
    &= \left(\frac{R_{H}}{\ell_{P}}\right)^2 \frac{d^2}{(2\gamma_{+})^2 \bar{a}_{0}^6}\left[\frac{(2\gamma_{+})^{\gamma_{+}}(2\gamma_{-})^{\gamma_{-}}}{C_{2}}\right]^{\frac{1}{2\gamma_{+}\gamma_{-}}}.
    \label{eq:omega-d} 
\end{align}

Combining equations \eqref{eq:C2} and \eqref{eq:omega-d}, we find a condition that $a_{+}$ and $\epsilon_{+}$ must satisfy:

\begin{align}
    \nonumber \frac{a_{+}}{\epsilon_{+}^{\gamma_{+}/3}} &= \frac{1}{(a_{b}\mathcal{X}_{b})^{(4\gamma_{+}\gamma_{-} - 1)}} \times \\
    & \times \left[\left(\frac{R_{H}}{\ell_{P}}\right)^{4\gamma_{+}\gamma_{-}}\frac{d^{4\gamma_{+}\gamma_{-}}2^{\gamma_{-}}(2\gamma_{+})^{\gamma_{+}}(2\gamma_{-})^{\gamma_{-}}}{\Omega_{d}^{2\gamma_{+}\gamma_{-}}}\right]^{1/3}.
    \label{eq:matching-1}
\end{align}
This condition determines the end of the quantum phase. As the quantum system evolves, we calculate the scale factor and $\epsilon$ in each step until the above matching condition. $\epsilon$ can be calculated within the quantum regime through the definition of $x$ in equation \eqref{eq:x-y}, the quantum planar system equations \eqref{eq:quantum-planar-1} and \eqref{eq:quantum-planar-2}, and the relation $x = 1-\epsilon_{+}$. From this point on, the classical regime starts with the matching values of $\alpha$ and $x$ as initial condition for the classical equation \eqref{eq:classical-equation}.  

\subsection{Matching with the dark energy scale}

There is a dark energy era in the expansion branch. Similarly to what we did in the previous section, we can choose $a_{0}$ to be the point where $\omega = -1$, which implies $x = 0$. From equations \eqref{eq:aH} and \eqref{eq:a}, we can write

\begin{equation}
    \frac{C_{1}}{\left(\frac{\lambda}{\sqrt{6}}\right)^{2}} = \Omega_{\Lambda}, \quad \text{and} \quad \frac{C_{2}}{\left(\frac{\lambda}{\sqrt{6}}\right)^{4\gamma_{+}\gamma_{-}}} = 1.
    \label{eq:C1C2-1}
\end{equation}

On the other hand, we can expand the same equations around the matching point $x_{-} = -1 + \epsilon_{-}$ and equate it to \eqref{eq:classical-limit-2} to obtain 

\begin{align}
    \nonumber C_{1} &= \left(\frac{R_{H}}{\ell_{P}}\right)^{2} d^2 \frac{1}{(2\gamma_{-})^2 \bar{a}_{0}^{6}}, \\
    C_{2} &= \frac{1}{2^{\gamma_{+}}\epsilon_{-}^{\gamma_{-}} (2\gamma_{-})^{4\gamma_{+}\gamma_{-}}}\left(\frac{a_{-}}{\bar{a}_{0}}\right)^3.
    \label{eq:C1C2-2}
\end{align}

The combination of equations \eqref{eq:C1C2-1} and \eqref{eq:C1C2-2} leads to the following match condition in the dark energy branch:

\begin{equation}
    \frac{a_{-}}{\epsilon_{-}^{\gamma_{-}/3}} = \left[\frac{R_{H}}{\ell_{P}} \frac{|d|2^{\gamma_{+}}(2\gamma_{-})^{(4\gamma_{+}\gamma_{-}-1)}}{\left(\frac{\lambda}{\sqrt{6}}\right)^{(1-4\gamma_{+}\gamma_{-})}\sqrt{\Omega_{\Lambda}}}\right]^{1/3}.
\end{equation}
As in the contracting phase, when the scale factor and the parameter $\epsilon$ reach the above condition, the quantum evolution in the expanding branch ends. The final state of the quantum phase will be the initial condition of the classical evolution.

\bibliography{references}

\end{document}